\documentclass[aip,reprint]{revtex4-1}

\draft 

\usepackage{graphicx}
\usepackage{dcolumn}
\usepackage{bm}

\usepackage[utf8]{inputenc}
\usepackage{selinput}
\usepackage[T1]{fontenc}
\usepackage{mathptmx}
\usepackage{etoolbox}
\usepackage{comment}
\usepackage{xcolor}
\usepackage{float}
\makeatletter
\def\@email#1#2{%
 \endgroup
 \patchcmd{\titleblock@produce}
  {\frontmatter@RRAPformat}
  {\frontmatter@RRAPformat{\produce@RRAP{*#1\href{mailto:#2}{#2}}}\frontmatter@RRAPformat}
  {}{}
}%
\makeatother
\begin{document}

\preprint{AIP/123-QED}

\title[Polariton lasing in AlGaN microring with GaN/AlGaN quantum wells]{Polariton lasing in AlGaN microring with GaN/AlGaN quantum wells}
\author{Anthonin Delphan}
\author{Maxim N. Makhonin}%
 \author{Tommi Isoniemi}
  \author{Paul M. Walker}
  \author{Maurice S. Skolnick}
 \author{Dmitry N. Krizhanovskii}
\affiliation{ Department of Physics and Astronomy, University of Sheffield, Sheffield, S3 7RH, UK } 

\author{Dmitry V. Skryabin}
\affiliation{Department of Physics, University of Bath, Claverton Down, Bath BA2 7AY, UK}

\author{Jean-Fran\c{c}ois Carlin}

\author{Nicolas Grandjean}

\author{Rapha\"{e}l Butt\'{e}}

\affiliation{Institute of Physics, \'{E}cole Polytechnique F\'{e}d\'{e}rale de Lausanne (EPFL), CH-1015 Lausanne, Switzerland}

\date{\today}

\begin{abstract}
 Microcavity polaritons are strongly interacting hybrid light-matter quasiparticles, which are promising for the development of novel light sources and active photonic devices. Here, we report polariton lasing in the UV spectral range in microring resonators based on GaN/AlGaN slab waveguides, with experiments carried out from 4 K up to room temperature. Stimulated polariton relaxation into multiple ring resonator modes is observed, which  exhibit threshold-like dependence of the emission intensity with pulse energy. The strong exciton-photon coupling regime is confirmed by the significant reduction of the free spectral range with energy and the blueshift of the exciton-like modes with increasing pulse energy. Importantly, the exciton emission shows no broadening with power, further confirming that lasing is observed at electron-hole densities well below the Mott transition. Overall, our work paves the way towards development of novel UV devices based on the high-speed slab waveguide polariton geometry operating up to room temperature with potential to be integrated into complex photonic circuits.
\end{abstract}

\maketitle

Microcavity polaritons are hybrid light-matter quasiparticles, arising from strong exciton-photon coupling in semiconductor photonic structures. They have attracted a significant attention in the last 15 years with a number of fundamental effects observed such as polariton condensation \cite{Kasprzak2006} and lasing \cite{bajoni2008}, superfluidity \cite{Amo2009}, solitons \cite{Sich2016}, polariton blockade \cite{Delteil2019} and single polariton nonlinearity \cite{Kuriakose2022} to name just a few, which are enabled by giant polariton interactions. Interacting polaritons are highly promising for the development of novel quantum light sources, photonic nonlinear simulators \cite{Amo2016}, logic gates and quantum optical signal processing \cite{Ghosh2020}. 

Polariton lasing, the coherent light emission from polariton condensates, provides several advantages over standard photonic lasing, including operation without population inversion with a threshold lower than that in conventional semiconductor lasers \cite{imamoglu1996}. It has been demonstrated via optical \cite{christopoulos2007,bajoni2008} and electrical pumping \cite{schneider2013}. Room temperature (RT) operation has been reported in structures based on wide bandgap semiconductors \cite{christopoulos2007,Feng2013,christmann2008} enabled by their large exciton binding energy. Polariton lasing has been mainly studied in planar semiconductor microcavities (MCs) made of two Bragg mirrors, which are challenging to fabricate. On the other hand polaritons have also been investigated in the slab waveguide (WG) geometry \cite{Walker2013,Ciers2017}, where photonic confinement in the vertical direction arises from total internal reflection (TIR). The main advantages of the WG geometry over MCs are low disorder due to thinner and simpler structures and the high polariton speed, enabling long propagation distances up to several 100s of $\mu$m, which makes this system favourable for the development of integrated polariton circuits. A number of nonlinear effects arising from giant optical Kerr-like polariton nonlinearity, such as dark and bright solitons \cite{Walker2015, Walker2017}, continuum generation \cite{Walker2019} and ultrafast pulse modulation \cite{DiPaola2021}, have been reported in the WG polariton platform.

III-nitride based polaritons are of particular interest since they enable coherent emission and low threshold ultrafast nonlinear optical modulation in the UV spectral range and can operate at RT \cite{DiPaola2021}, with many potential applications including studies of chemical reactions, coherent Raman spectroscopy, and manipulation of trapped ions. UV polariton lasing in III-nitride WG devices has been reported only recently in GaN ridge resonators up to 150 K, likely limited by the smaller exciton binding energy in bulk GaN \cite{Souissi2022} compared to that in quantum heterostructures.

In this paper, we report polariton lasing in microring resonators fabricated from GaN/AlGaN quantum well (QW) slab WGs with operation in a wide temperature range  from 4 to 300 K enabled by the large exciton binding energy in the GaN QWs ($\sim 48$ meV) and a lower surface state recombination velocity than their III-arsenide counterparts \cite{Boroditsky2000}. Polariton lasing from the multiple ring resonator modes is revealed by the threshold-like dependence of the emission intensity with increasing pump pulse energy. The free spectral range (FSR) between the modes reduces as the polariton energy approaches that of the exciton due to the strong dependence of the polariton group velocity on the photonic fraction of the polaritons, a clear signature of strong exciton-photon coupling. Furthermore, a strong blueshift of the lower polariton exciton-like states with increasing pulse energy is observed due to polariton interactions. 

Apart from the development of coherent UV polariton sources, our demonstration of microring polariton resonators also paves the way towards further applications in integrated polariton circuits (e.g. filtering and directional coupling) and studies of low threshold generation of frequency combs and Kerr solitons.

Our microring resonators are formed by etching an AlGaN planar slab WG containing multiple GaN QWs \cite{Ciers2017} which was grown by metal-organic vapour phase epitaxy on free-standing GaN substrate. Propagating polaritons have been demonstrated in similar unetched planar WGs with a Rabi-splitting  of $\sim$ 60 meV \cite{Ciers2017}. The confinement induced by the ring geometry is expected to lead to discrete clockwise and counterclockwise ring polariton modes. The samples are fabricated by means of e-beam lithography and reactive ion etching. 

After etching, the total height of the structures amounts to $\sim$ 315 nm (Fig. 1a). Rings of different radii ($R$) and widths ($t$) have been etched on the sample. The radius is taken from the centre to the mid-point of the ring. Here we studied microrings with $R$ = 3, 4, 6, 8 $\mu$m and $t$ = 2 $\mu$m. A scanning electron microscopy (SEM) image of a typical microring ($R$ = 3 $\mu$m, $t$ = 1 $\mu$m) is shown in Fig. 1b (see also Fig. S1 in the supplemental material (SM)). \par 

\begin{figure}
    \centering
    \includegraphics[width=85mm]{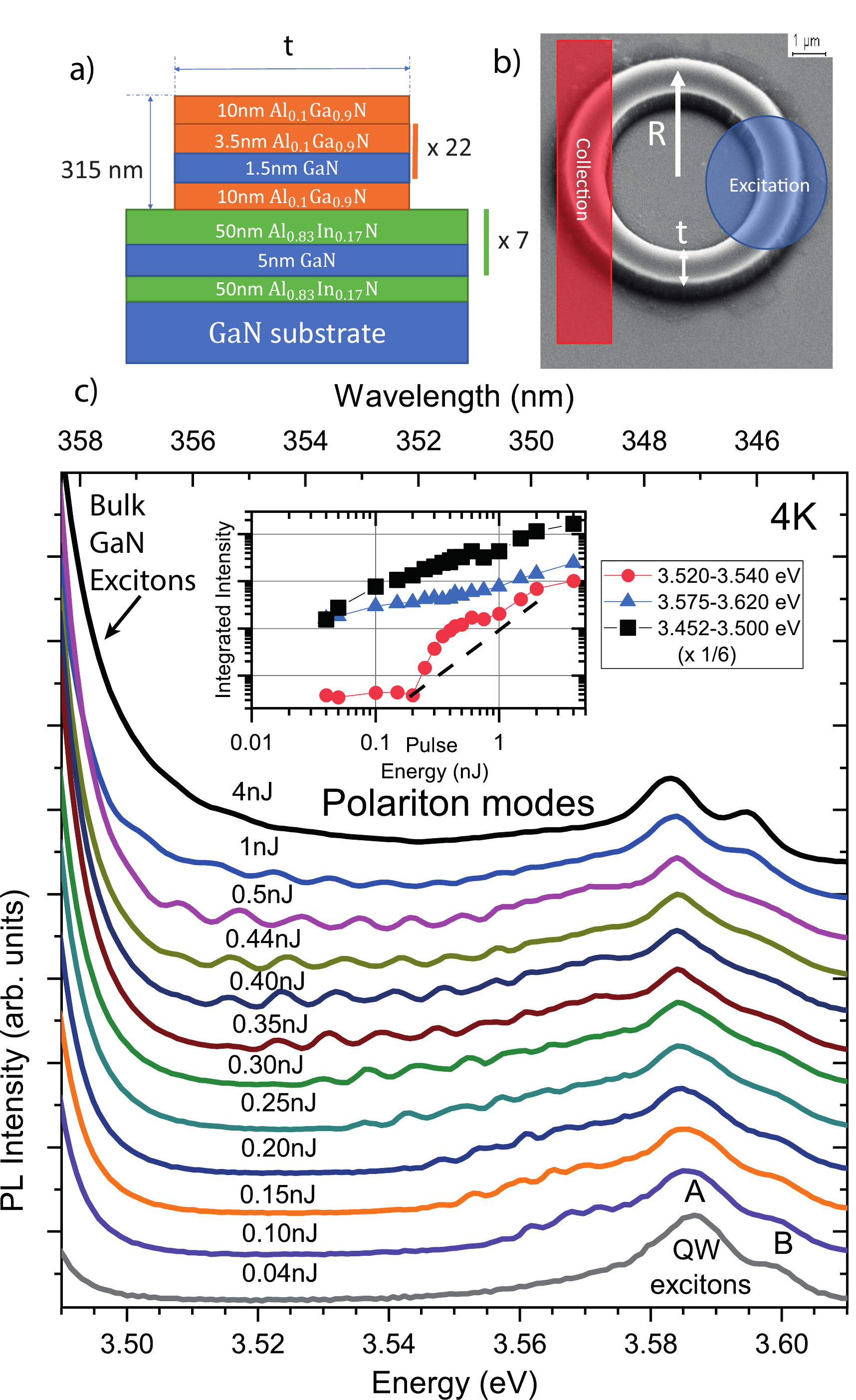}
    \caption{a) Schematic of the ring resonator design. b) SEM image of a ring resonator at a tilted angle of 30$^\circ$ from the surface normal, $R$ = 3 $\mu$m, $t$ = 1 $\mu$m with overlaid excitation and collection areas. c) PL spectra (linear scale) of an AlGaN resonator with $t$ = 2 $\mu$m and $R$ = 4 $\mu$m collected at $T$ = 4 K for pulse energies ranging from 0.04 nJ to 4 nJ. The PL intensity is normalised to the intensity of the QW A exciton: the signal obtained from the exciton is divided by a different factor for each power so that it is equal to unity for all powers. The spectra are then shifted for clarity by half-unity. See also SM section 2. Inset: Integrated intensities (log-log scale) for the region with modes on which background contribution is negligible (red),
    the region corresponding to the QW exciton (blue) and the region corresponding to the bulk GaN exciton (black). The black dashed line represents a quadratic increase.}
    \label{fig:power_dependency}
\end{figure}

We study polariton lasing over a wide range of temperatures ($T=$ 4-300 K) by using a continuous flow liquid helium cryostat. A standard microphotoluminescence ($\mu$-PL) setup allows us to excite and collect light emission at different spots on the sample in a backscattering configuration (see Fig. 1b). Pulsed laser excitation is performed with a frequency-quadrupled optical parametric amplifier (TOPAS) pumped by a Ti:sapphire regenerative amplifier. The excitation pulses were centred around 320 nm and had a duration of 100 fs and a repetition rate of 1 kHz. The PL signal collected by a microscope objective (NA = 0.39) is sent to a spectrometer with a resolution of 0.16 nm while being integrated over the entire rectangular entrance slit of the monochromator.  

The optical resonances in the ring structures occur when the circumference of the ring is an integer multiple ($n$) of the polariton wavelength. This condition can be written as $k(E)2\pi r_{\text{eff}} =2\pi n$, where $k(E)$ is the polariton wavenumber at energy $E$ and $r_{\text{eff}}$ is the effective radius at which the mode propagates. In addition, the finite thickness of the ring may give rise to quantisation of polariton waves along the radial direction and formation of transverse modes characterised by the integer quantum number $m$. We calculated the optical modes of the circular waveguides using the Lumerical MODE finite difference eigenmode (FDE) solver (see supplementary discussion 4).

To demonstrate polariton lasing, we carried out a PL pulse energy dependence study on the $R = 4$ $\mu$m, $t = 2$ $\mu$m resonator with pulse energies varying from 0.04 to 4 nJ. 
The emission spectra for this entire excitation range at $T=$ 4 K are shown in Fig. \ref{fig:power_dependency}c. At low pulse energies the spectra consist of excitonic peaks at $\sim$3.587 eV and $\sim$3.598 eV associated with QW A and B excitons as previously reported in studies led on similar wafer samples \cite{Ciers2017}.  Even though the detection and excitation regions are spatially separated, exciton emission in the detection region may arise from the propagation of high velocity upper polaritons away from the excitation spot with their subsequent relaxation into the lower energy low momenta exciton states. Upon increasing pulse energy, narrow full width at half maximum (FWHM) $\sim$3 meV) modes start to appear abruptly at energies below the QW exciton peaks, firstly closer to the QW exciton emission, and then at lower energies down to 3.501 eV between the QW excitonic peaks and the bulk GaN exciton emission (centred at $\sim$3.46 eV) originating from the substrate. 

As we argue below, the appearance of these modes with increasing pulse energy is associated with stimulated polariton scattering into the optical ring resonator states which are in the strong coupling regime with the QW excitons. The bulk GaN excitons are isolated from the ring structure by the cladding layers and thus are only weakly coupled to the resonator modes. Importantly, our polariton system is nonequilibrium: the multiple modes (co-existing condensates) become macroscopically occupied due to the dynamical equilibrium between gain and dissipation channels \cite{Kasprzak2008, Krizhanovskii2009}. At 4 K, relaxation to the polariton states occurs mainly through exciton-exciton and exciton-polariton scattering, since at low temperature phonon-assisted scattering is inefficient \cite{Ciers2020}. Initially, at pulse energies just above threshold, scattering to the exciton-like polariton states whose energy is close to the QW exciton level is more efficient. With further increase in pulse energy scattering to more photon-like states increases \cite{Tartakovskii2000} leading to polariton lasing from the lower energy ring states (see Fig. \ref{fig:power_dependency}c).
The macroscopical occupation of the polariton modes is confirmed by the superlinear (threshold-like) increase of the  mode emission intensity (integrated in the energy interval where the background from GaN and QW excitons is negligible) with pulse energy, whose dependence is much faster than quadratic, as shown in the inset of Fig. \ref{fig:power_dependency}c. If the filling factor of the polariton modes was less than unity it is expected that enhanced relaxation due to interparticle scattering would lead only to a quadratic dependence. For comparison, we also show in the inset of Fig. \ref{fig:power_dependency}c that the integrated emission intensity exhibits a linear or slightly superlinear power dependence in the energy range where QW and GaN backgrounds dominate, respectively.

We note that in our case the modes are confined in the vertical direction due to TIR and the emission is likely observed due to Rayleigh scattering on fabrication imperfections of the ring resonators. 

Below threshold, the polariton emission from the ring resonator modes is too weak to be detected because the latter is guided in the microring resonator plane. It is only above threshold that the scattered light from the lasing modes becomes sufficiently strong and comparable to the excitonic background to be coupled to our microscope objective. 

\par
\begin{figure}
    \centering
    \includegraphics[width=80mm]{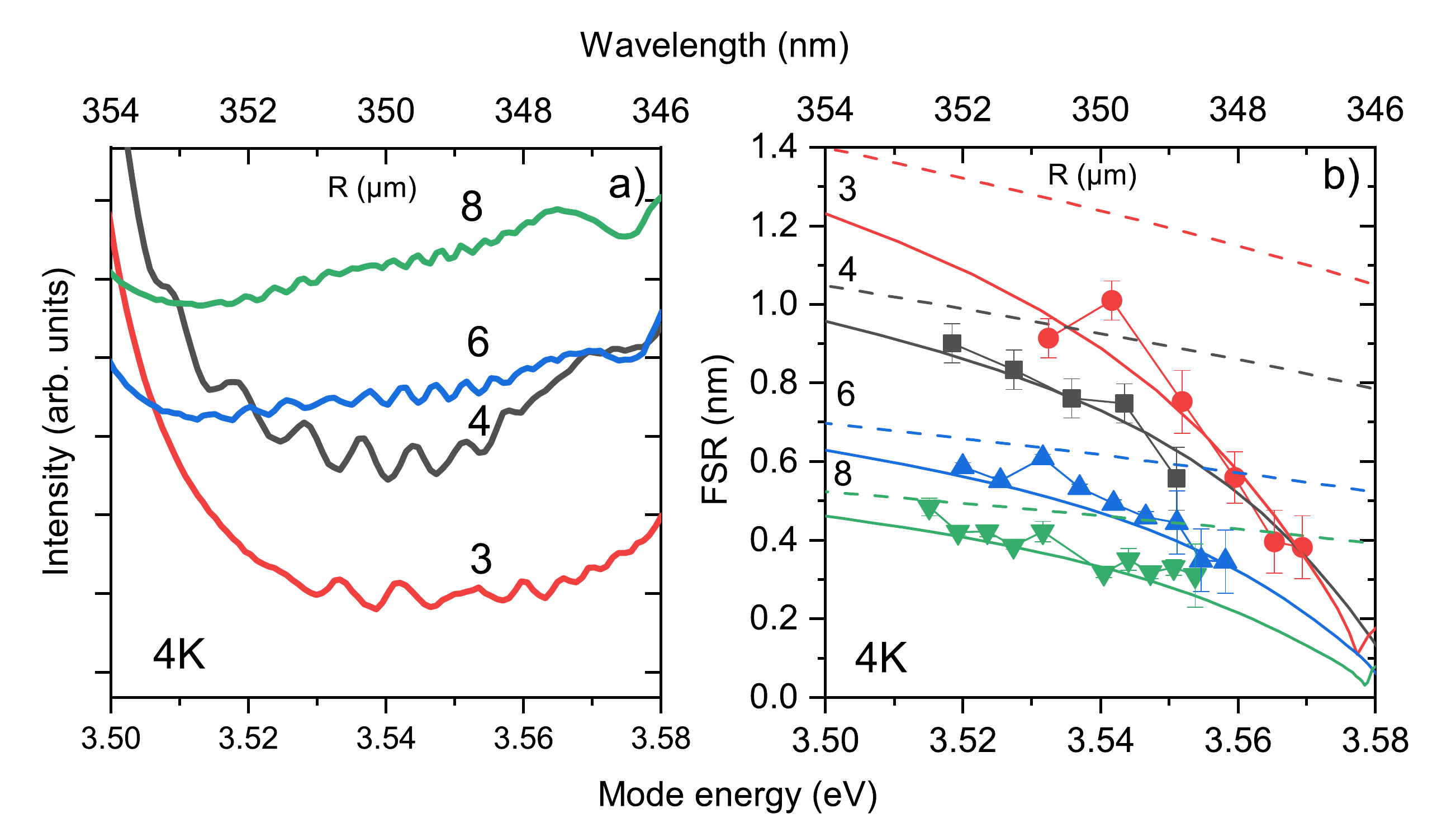}
    \caption{a) PL spectra in linear scale of microring resonators measured for different radii (3, 4 ,6, and 8 $\mu$m), with a 2 $\mu$m width, at pulse energy above threshold (0.4 nJ) taken at 4 K. b) Free spectral range of the same rings. The solid lines are a fit according to Eq. (1) accounting for the excitonic content of the guided modes. The dashed lines show the results of a purely photonic model. Simulation details and FSR data recorded at 300 K are shown in the SM (sections 4 and 5).}
    \label{fig:FSR}
\end{figure}

The FSR of the ring resonator modes is given by the following equation: 
\par 
\begin{equation}\label{eq:FSR}
    \text{FSR} = \frac{\hbar v_{G}\left(E\right)}{r_{\text{eff}}},
\end{equation}
\par
where $\hbar$ is the reduced Planck's constant and $v_{G}\left(E\right) $ is the (energy dependent) group velocity of planar slab WG polariton modes at energy $E$, and $r_{\mathrm{eff}}$ is the radius around which the lasing modes propagate. In the strong coupling regime $v_{G}\left(E\right)$ and hence the FSR are expected to decrease strongly with increasing $E$ as the lower polariton dispersion curves strongly towards the exciton level (see supplementary discussion 5). 
We investigated the stimulated emission from ring resonators of different radii at 4 K over the wavelength range 346-354 nm, with the raw data shown in Fig. \ref{fig:FSR}.a). The FSR vs. energy is summarised in  Fig. \ref{fig:FSR}.b). The FSR increases with decreasing ring radius $R$, as expected from Eqn.~\ref{eq:FSR}. The FSR also decreases with increasing energy. For $R=3$ and 4 $\mu$m  the FSR decreases drastically by a factor of 2-3 as the energy approaches the exciton level, which is a strong confirmation that the observed modes are in the strong coupling regime. Some trend of decreasing FSR with energy is also visible for $R$ = 6 and 8 $\mu$m, although the dependence is more noisy for $R$ = 8 $\mu$m since the FSR becomes comparable to the spectral resolution. The theoretical FSR for purely photonic modes was calculated using FDE as described above and the FSR for polaritons was calculated using the photonic values and a standard coupled oscillator model (see supplementary section 5). The solid lines in Fig. \ref{fig:FSR}.b show the theoretical polariton FSR and their curvature is in good agreement with the experiment. The purely photonic FSR theory curves (dashed lines) also show decreasing FSR with energy which occurs mainly due to the energy dependence of the material refractive indexes. However, this curvature is much shallower than that seen in the experiment. In summary, the strong curvature of the FSR vs. energy cannot be explained without invoking a strong coupling picture, thus confirming that the system remains strongly coupled while in the lasing regime. Further detail is given in supplementary section 5.

\begin{figure}
    \centering
    \includegraphics[width=80mm]{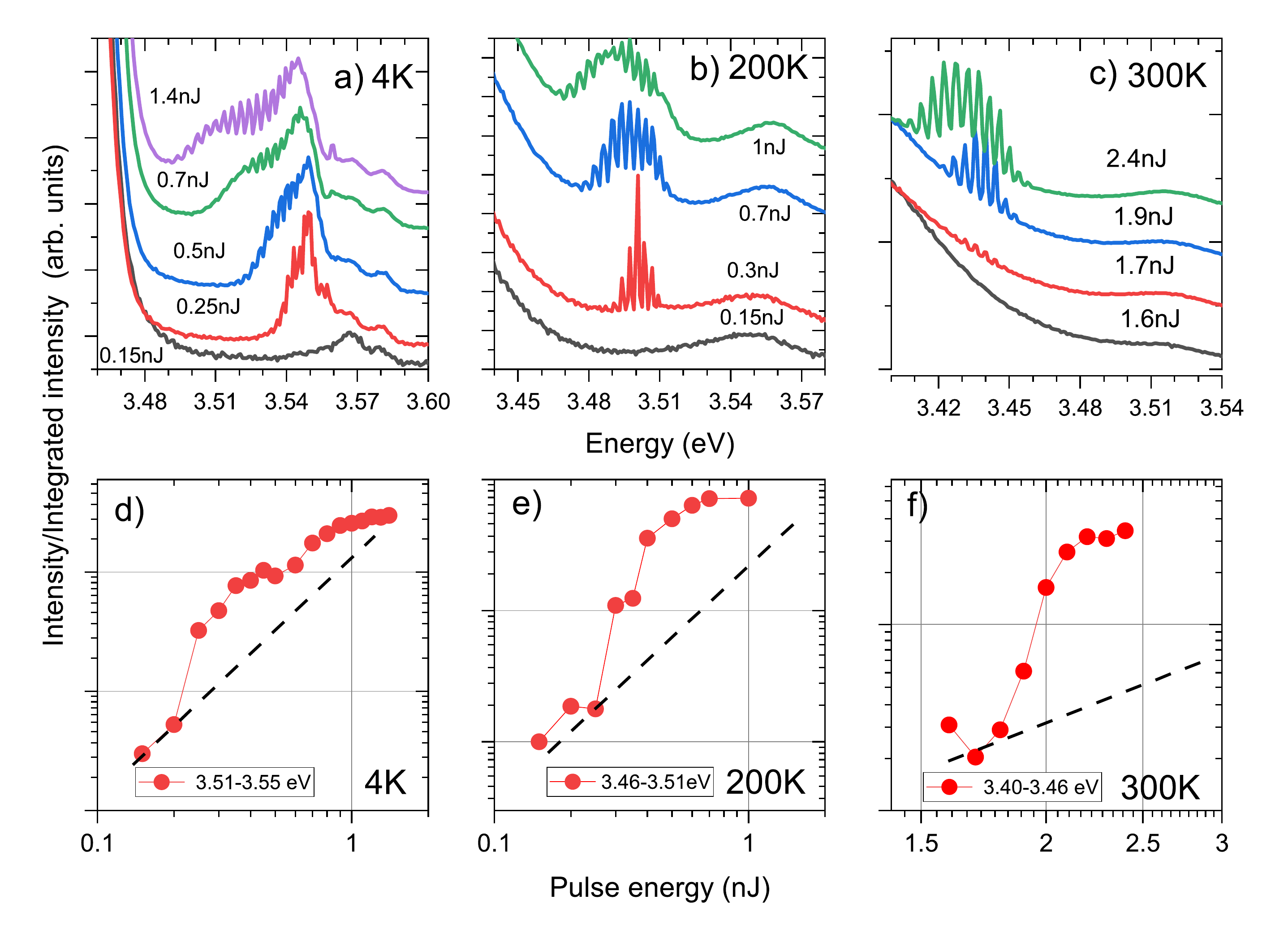}
    \caption{(a), (b), (c) PL spectra for the 8 $\mu$m radius, 2 $\mu$m wide ring at different pulse energies (in nJ) labeled in the figures for $T=$ 4, 200, and 300 K, respectively. Each spectrum is normalised to the QW A exciton (3.566 eV at 4 K, 3.547 eV at 200 K, 3.511 at 300 K) and then shifted for clarity. Non-shifted values and an alternative presentation are given in SM section 2. (d), (e), (f) Integrated intensity of the modes within given energy boundaries plotted in log-log scale for $T=$ 4, 200, and 300 K, respectively. The black dashed line represents a quadratic increase. 
    }
    \label{fig:threshold}
\end{figure}

Furthermore, we perform  measurements on an $R$ = 8 $\mu$m microring over a wider temperature range up to 300 K. The PL spectra for different pulse energies are displayed in Figs. \ref{fig:threshold}a,b,c for $T=$ 4, 200 and 300 K, respectively. At all temperatures the bunch of narrow ring resonator modes associated with polariton lasing  appears below the QW A exciton energy level with increasing pulse energy. As in the case of $R$ = 4 $\mu$m the emission intensities integrated in the energy range where modes are the most visible, demonstrate a clear threshold-like behaviour with a faster than quadratic dependence in Figs. \ref{fig:threshold}d,e,f at all temperatures, indicating macroscopic occupation and lasing of several modes. By contrast, the bulk GaN and QW excitonic peaks behave linearly (see SM section 7).
As the data taken at 200 and 300 K sits on a strong incoherent PL background originating from the bulk GaN and the QW exciton emission peaks, this background has been subtracted. 
The resulting PL spectra are available in the SM (section 6). In the supplementary material (SM section 5), we plot FSR measurements at 300 K and compare them with the theorerical simulations, which support the strong exciton-photon coupling at room temperature.

Whilst with increasing temperature both the exciton and the polariton emission shift to lower energy due to band gap reduction, it is also observed that the onset of lasing occurs in more photon-like polariton states at higher $T$. This occurs because a) the losses of the exciton-like states with energies closer to the exciton level increase due to phonon scattering and b) at elevated temperature the polaritons can relax to more photon-like states by exciton-phonon as well as exciton-exciton and exciton-polariton scattering~\cite{Ciers2020,Tartakovskii2000,Savvidis2002}. The non-radiative processes and polariton losses increasing with temperature also lead to the increase of the lasing threshold  from 0.25 nJ at 4 K to 0.35 nJ and approximately 1.8 nJ at 200 K and 300 K, respectively. Interestingly, however, this overall increase in the polariton lasing threshold with temperature remains within a factor about seven. Such a variation is far smaller than that reported recently for ridge waveguide polariton lasers made from bulk GaN, where the drastic increase in the lasing threshold by more than two orders of magnitude from 70 K to 220 K was incompatible with a polaritonic picture that could hold from cryogenic to room temperature \cite{Souissi2022}. This reduced sensitivity for our sample is again most likely stemming from the quantum heterostructure nature of our polariton gain medium that leads to stabler excitons. We are also able to explore a significant range of pumping conditions above threshold with pulse energy values nearly up to an order of magnitude larger than that at threshold at 4 K for the 8 $\mu$m ring resonator (Figs. \ref{fig:threshold}d,e,f) and even beyond one order of magnitude for the 4 $\mu$m ring resonator (Fig. \ref{fig:power_dependency}c and corresponding inset).

\begin{figure}[H]
    \centering
    \includegraphics[width=90mm]{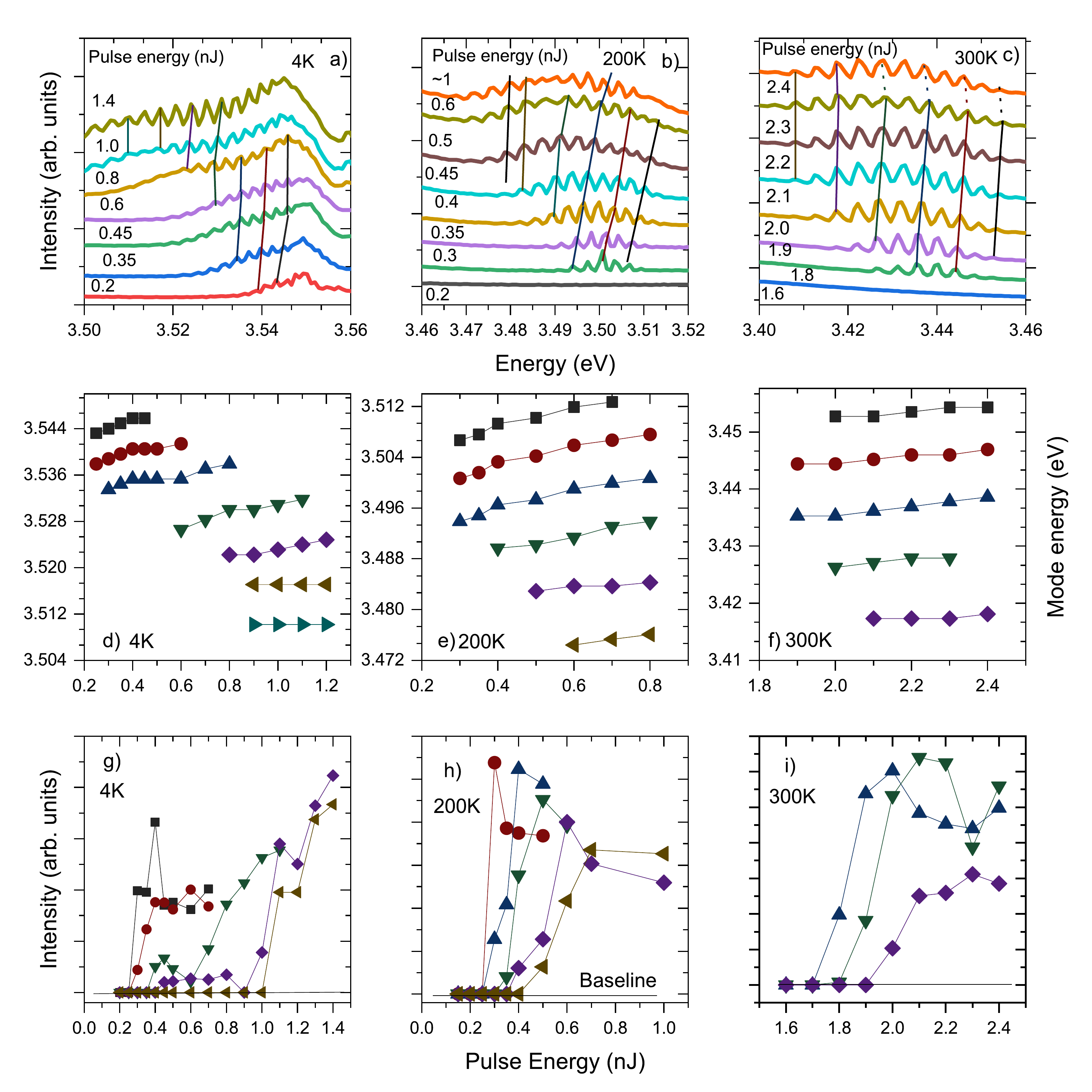}
    \caption{(a), (b), (c) PL spectra of the 8 $\mu$m ring resonator taken at $T=$ 4, 200, and 300 K, respectively. The solid lines act as guides for the eye to indicate the shift of selected modes with pulse energy. The dashed lines represent points where the shift is less visible. Data is obtained from Figs. \ref{fig:threshold}. (d), (e), (f) Mode energy versus pulse energy of selected modes, identified by the line colour, at 4, 200, and 300 K, respectively. (g), (h), (i) The peak mode intensity versus pulse energy of selected polariton modes, identified by their line colour, at 4, 200, and 300 K, respectively. 
    }
    \label{fig:blueshifts}
\end{figure}

The excitonic component of the polariton wavefunction leads to strong polariton-polariton  interactions responsible for the blueshift of polariton resonances with increasing density.  
In Figs. \ref{fig:blueshifts}a,b,c we plot expanded spectra for different pulse energies, tracing the peak position of each of the lasing modes. The extracted peak positions of the polariton modes are plotted in Figs. \ref{fig:blueshifts}d,e,f, respectively. As expected the higher energy exciton-like states exhibit stronger blueshift with pulse energy than the lower energy photon-like polaritons. At 4 and 200 K an energy blueshift up to 6-7 meV is observed for exciton-like modes, whereas at 300 K the observed shifts are much less, $\sim1-2$ meV, due to the decreased excitonic content of the lasing modes and increased thermal effects, which may lead to polariton redshift counterbalancing the effect of interactions. Note that only inter- and intra-mode polariton-polariton interactions are responsible for the polariton blueshifts, whereas the interaction with the higher energy exciton reservoir does not play a role since its density is expected to be pinned above the polariton lasing threshold \cite{Wouters2007}. The evolution of the peak intensity of the modes with pulse energy is finally given in Fig. \ref{fig:blueshifts}g,h,i. The peak intensity is given by the maximum intensity of the mode minus the intensity at the base of the peak. We can see that the modes examined here have a different threshold-like increase at a given pulse energy, and then stagnate or decrease as more modes, and hence mode competition, come into play. Similar strong blueshifts are seen in the 4 $\mu$m ring at 4 K (see SM section 8). Finally, we note that interparticle interactions and mode competition may determine the linewidth (FWHM$\sim$ 2-3 meV) of the lasing modes above threshold (see SM section 3).

\par
Importantly, the fact that the exciton emission line, detected either in the area of the pump spot or on the opposite side of the ring, (Figs. \ref{fig:power_dependency} and \ref{fig:threshold}) does \textit{not} broaden with pulse energy and shows no or small energy blueshift ($\sim$10 meV) confirms that there is a limited screening of the built-in electric field and that the created electron-hole density in each QW is well below the Mott density ($\sim10^{12}$ cm$^{-2}$) \cite{Rossbach2014}. Indeed, above the Mott density, the emission is expected to originate from an electron-hole plasma with a high energy emission tail extending by 50-60 meV from the exciton peak maxima \cite{Rossbach2014}. This observation is another confirmation that the nonlinear emission is associated with polariton lasing.  
\par
In conclusion, we report UV polariton lasing from multiple modes in microring resonators fabricated from AlGaN planar waveguides with embedded GaN quantum wells at temperatures up to 300 K.  The micro-structured polariton system we present has the potential to be used to study modelocking of polaritons into a sequence of short pulses, generation of UV soliton trains and frequency combs supported by giant polariton Kerr nonlinearity. Polariton modelocking was demonstrated numerically for resonantly pumped microring resonators \cite{opex}. Beyond this, the non-resonant pump like the one used in this work and a combination of both resonant and non-resonant pumping offer several avenues for further research into the complex interplay between turbulence and modelocking~\cite{Piccardo2020}. Additional opportunities should arise from the coupling between two and more rings, which includes combining space topology and modelocking~\cite{segev}, an aspect not easily within reach with the linear WG geometry. Overall, our work has the potential to be a significant step forward for the development of compact
active nonlinear polariton devices operating at RT. 

\section*{Supplementary material}
See supplemental document for supporting information concerning experimental methods and simulation models for polariton dispersion.

\section*{Author Contribution} 

\textbf{Anthonin Delphan}: Data Curation, Formal Analysis, Investigation, Software, Visualisation, Conceptualisation, Methodology, Validation, Writing - Original Draft \\
\textbf{Maxim N. Makhonin}: Data Curation, Formal Analysis, Investigation, Software, Visualisation, Conceptualisation, Supervision, Methodology, Validation, Writing - Original Draft \\
\textbf{Tommi Isoniemi}: Resources, Investigation, Methodology, Writing - Review \& Editing \\
\textbf{Paul M. Walker}: Data Curation, Formal Analysis, Software, Visualisation, Conceptualisation, Methodology, Writing - Review \& Editing \\
\textbf{Maurice S. Skolnick}: Conceptualisation, Methodology, Writing - Review \& Editing \\
\textbf{Dmitry N. Krizhanovskii}: Conceptualisation, Funding acquisition, Methodology, Project Administration, Supervision, Writing - Review \& Editing \\
\textbf{Dmitry V. Skryabin}: Methodology, Writing - Review \& Editing \\
\textbf{Jean-Fran\c{c}ois Carlin}: Methodology, Resources, Writing - Review \& Editing \\
\textbf{Nicolas Grandjean}: Methodology, Resources, Writing - Review \& Editing \\
\textbf{Rapha\"{e}l Butt\'{e}}: Conceptualisation, Methodology, Project Administration, Supervision, Writing - Review \& Editing \\

\begin{acknowledgments}
We acknowledge UK EPSRC grants EP/V026496/1, EP/S030751/1 and EP/R007977/1. Anthonin Delphan would like to thank Christopher Vickers and Thomas Ball for their help with the cryogenics setup.
\end{acknowledgments}

\section*{Conflicts of interests}
The authors declare no conflicts of interest.

\section*{Data Availability Statement}

\begin{center}
\renewcommand\arraystretch{1.2}
\begin{tabular}{| >{\raggedright\arraybackslash}p{0.3\linewidth} | >{\raggedright\arraybackslash}p{0.65\linewidth} |}
\hline
\textbf{AVAILABILITY OF DATA} \& \textbf{STATEMENT OF DATA AVAILABILITY}\\  
\hline
Data available on request from the authors
&
The data that support the findings of this study are available from the corresponding author upon reasonable request.
\\ \hline

\end{tabular}
\end{center}

\nocite{*}
\bibliography{aipsamp}

\section*{Polariton lasing in AlGaN microring with GaN/AlGaN quantum wells - Supplemental Material}
\subsection{Growth details and fabrication process}

The sample was grown using metal organic vapour phase epitaxy on high-quality low threading dislocation density ($\sim$10$^{6}$ cm$^{-2}$) \textit{c}-plane free-standing GaN substrate. The sample structure consists of a 130-nm-thick active region with 22 embedded GaN/Al$_{0.1}$Ga$_{0.9}$N (1.5 nm/3.5 nm) quantum wells (QWs) grown on top of a 400-nm-thick Al$_{0.83}$In$_{0.17}$N cladding lattice-matched to GaN. It is essentially a replica of the samples in which waveguided polaritons were reported for the first time in III-nitride slab waveguides [1] and for which we subsequently demonstrated ultrafast-nonlinear ultraviolet pulse modulation in the strong coupling regime up to room temperature [2].
For electron beam lithography (EBL) a layer of the photoresist CSAR-62 [3] with the adhesion enhancer hexamethyldisilazane was spinned on the sample. The soft mask has a thickness of 1130 nm. The resist was exposed with a 50 keV electron beam using a Raith Voyager EBL system, using a base dose of 94 $\mu$C/cm$^2$ together with a proximity dose correction. The mask was developed in xylene for 60 s at 23 $^\circ$C and rinsed in isopropanol. After development, inductively-coupled plasma reactive ion etching (ICP-RIE) was used to etch the patterns through the core and partly into the cladding. The dry etch process uses a RF power of 80 W, ICP power of 450 W and 4 mTorr pressure. The gas flows were 1.5 sccm for SiCl$_4$, 15 sccm for Cl$_2$ and 4 sccm for Ar. The etching time was 10.75 min with the etching rates of 67 nm/min for the resist and 30 nm/min for GaN. After etching the resist was removed with 4 min O$_2$ plasma ashing with 100 W power, submerging the sample in heated Microposit resist remover 1165 for 4 min and rinsing it in isopropanol for 2 min. \par
\begin{figure}[H]
    \centering
    \includegraphics[width=80mm]{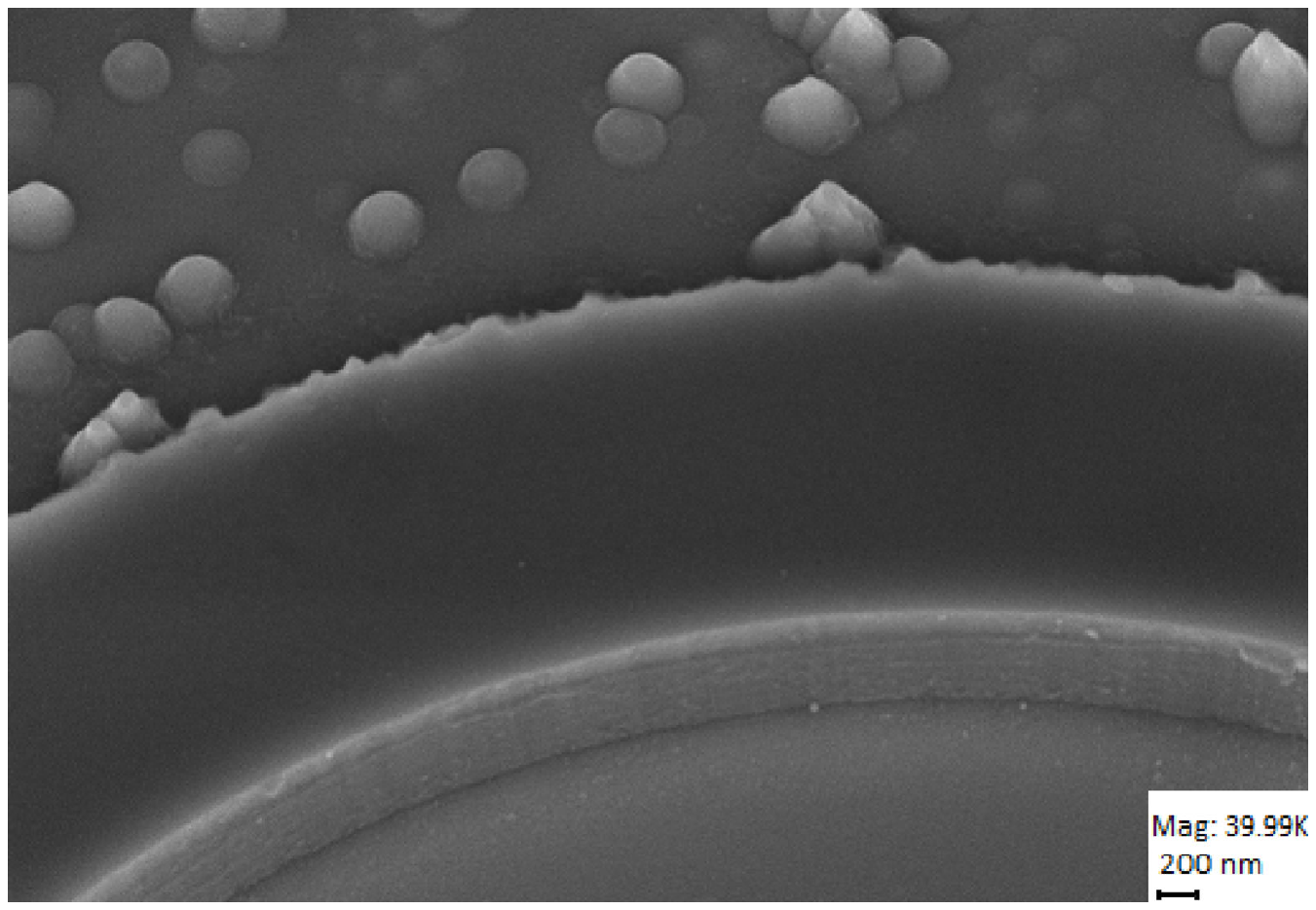}
    \caption{Close-up view of a ring resonator obtained with SEM imaging.}
    \label{fig:top}
\end{figure}

A scanning electron microscopy (SEM) image of a ring resonator is shown in Fig. \ref{fig:top} that illustrates the very good flatness of the top ring. The small droplets on the outer side of the ring are from unetched photoresist and are not believed to have any impact on the optical measurements. 

\subsection{Normalisation of Photoluminescence Spectra}

In the main text, we present in Fig. 1 the photoluminescence (PL) spectra of the 4 $\mu$m ring at different pulse energies. We also give PL spectra of the 8 $\mu$m ring in Fig. 3 for different pulse energies and temperatures. These spectra are normalised to the PL intensity of the QW A exciton, according to equation \ref{eq:norm}: 
\begin{equation} \label{eq:norm}
    \mathrm{PL}_N (\mathcal{E}) = \frac{\mathrm{PL}(\mathcal{E})}{\mathrm{PL}(\mathcal{E}_X)},
\end{equation}
where $\mathcal{E}$ is the energy,  $\mathrm{PL}$ is the raw photoluminescence data, $\mathcal{E}_X$ is the energy of the QW A exciton resonance, and $\mathrm{PL}_N$ is the resulting normalised spectrum. We observe that the function $\mathrm{PL}_N$ is almost constant for the QW peaks and the bulk GaN peaks across all pulse energies, which shows that both follow a similar pulse energy dependency (shown in section 6 to be quasi-linear), but varies significantly in the region where the modes appear. This shows that the polariton modes follow a non-linear pulse energy dependency.
\par
The PL spectra are all shifted vertically for clarity. The value of the shifts is as follows:
\begin{itemize}
    \item for the 4 $\mu$m ring at 4 K, the shift is equal to 0.5 for all powers;
    \item for the 8 $\mu$m ring at 4 K, the shift is equal to 0.7 for 0.25 nJ, then 1.5 for 0.5 nJ, 2 for 0.7 nJ, and finally 1 for 1.4 nJ;
    \item for the 8 $\mu$m ring at 200 K, the shift is equal to 1 for 0.3 nJ, then 2.5 for 0.7 nJ, and finally 1.5 for 1 nJ;
    \item for the 8 $\mu$m ring at 300 K, the shift is equal to 3 for all pulse energies.
\end{itemize}

\begin{figure}[H]
    \centering
    \includegraphics[width=80mm]{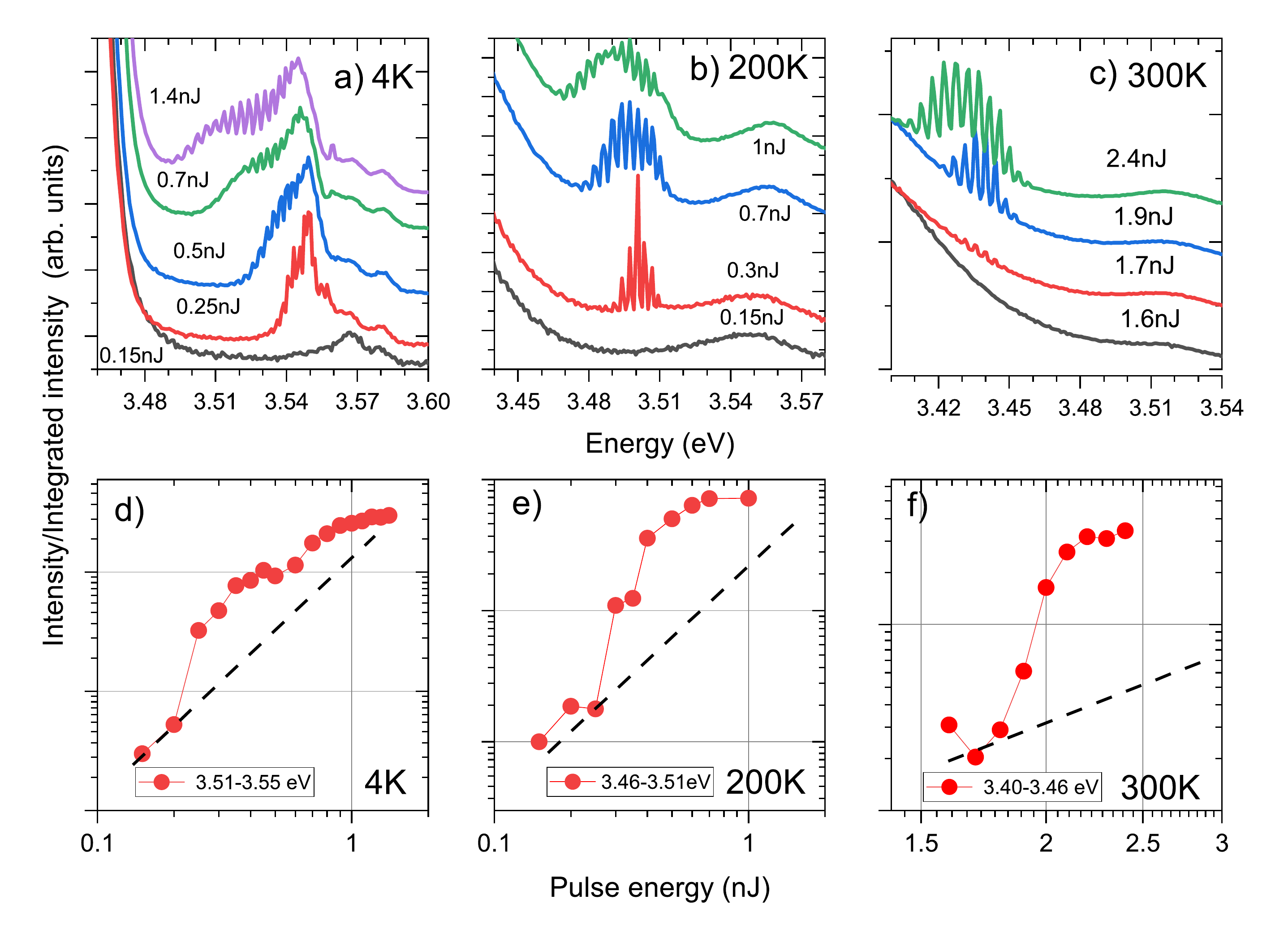}
    \caption{(a), (b), (c) PL spectra for the 8 $\mu$m radius, 2 $\mu$m wide ring at different pulse energies (in nJ) labeled in the figures for $T=$ 4, 200, and 300 K, respectively. Non-normalised data.}
    \label{fig:linear_scale}
\end{figure}

The interest of this representation is to emphasize the evolution in relative contrast between the polariton modes and the neighbouring excitons. In Fig. \ref{fig:linear_scale}, we show the non-shifted, non-normalised spectra, which show the evolution in absolute contrast.

\subsection{Polariton mode linewidth}

One of the distinctive signatures of lasing is a reduction of the linewidth of the cavity modes supporting lasing when crossing the threshold. For a single mode laser the linewidth of the lasing mode is predicted to decrease inversely proportional to the mode filling factor according to the celebrated Schawlow-Townes linewidth formula [4]. 
In our experiment, the scattered photoluminescence signal from the cavity modes is not observed below threshold, since it is too weak to efficiently contribute to the collected signal with our backscattering configuration as discussed in the main text. Above the polariton lasing threshold the peak intensities of the modes quickly saturate with power (see Fig. 4 of the main text) and no measurable reduction of the polariton mode linewidths is observed. The linewidth of the polariton modes is observed to be around 2-3 meV and is very likely limited by the polariton-polariton interactions, cross scattering and gain competition between the modes. Given that our system is pumped with a 1 kHz laser of very short (100 fs) pulses there is most likely a strong variation in the total exciton and polariton density (and hence in the modes peak positions) from pulse to pulse, which further leads to polariton mode broadening in the time-averaged spectra we measure.


\par
\par
\par

\subsection{Polariton dispersion and photonic loss of ring resonators}\label{sec:SI_dispersion}

\begin{figure}[H]
    \centering
    \includegraphics[width=80mm]{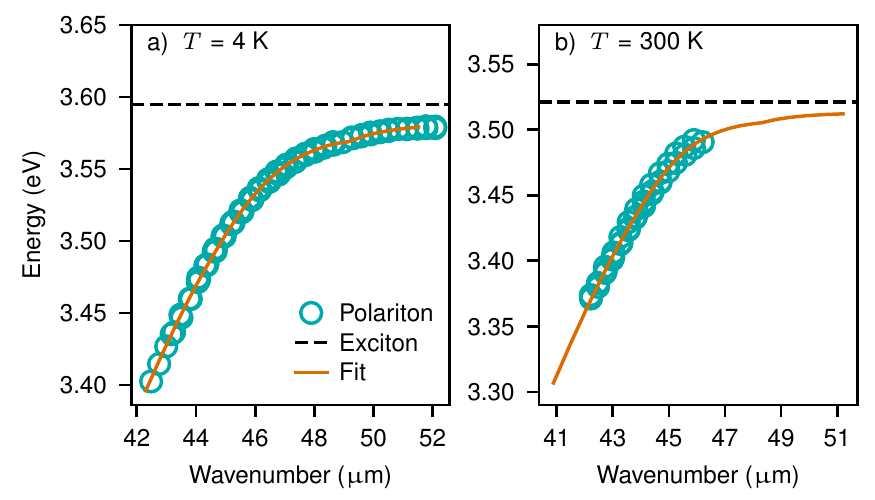}
    \caption{Dispersion of the unstructured planar waveguide. Experimental energy vs. wavenumber dispersion relation (points) and best fit of the coupled oscillator model (solid curve) for a) $T=4$ K and b) $T=300$ K.}
    \label{fig:planar_disp}
\end{figure}

The strong reduction in free spectral range (FSR) as the energy approaches that of the exciton resonance arises due to the strong dispersion of polaritons. This is illustrated in Fig.~\ref{fig:planar_disp} which shows the experimentally measured energy vs. wavenumber dispersion relation for the planar waveguide. The points give the experimental data while the solid curve gives the best fit of the polariton model. Panels a) and b) show the cases for two representative temperatures: 4 K and 300 K, respectively. The data and analysis are taken from Ref. [2] and are fully described in that work.

Exciton-polaritons (hereafter polaritons) are formed by hybridisation of photons and excitons. The polariton dispersion is obtained by first calculating the dispersion of the purely photonic modes of the planar waveguide structure using electromagnetic simulation, and then employing a coupled oscillator model to calculate the dispersion of the polaritons. The photons and excitons hybridise to form two polariton branches, the upper polaritons (UP) and lower polaritons (LP). The UP are not observed because of strong absorption at energies above the exciton resonance [1]. The dispersion of the LP given by the conventional coupled oscillator model is

\begin{equation}\label{eq:coupled oscillator}
    E_{\mathrm{LP}} = \left[E_{\mathrm{ph}}+E_{\mathrm{ex}} - \sqrt{\left(E_{\mathrm{ph}}-E_{\mathrm{ex}}\right)^2 + \Omega^2}\right]/2.
\end{equation}

Here $E_{\mathrm{LP}}$ is the lower polariton energy, $E_{\mathrm{ph}}$ and $E_{\mathrm{ex}}$ are the photon and exciton energies, respectively, and $\Omega$ is the coupling strength between photon and exciton, known as the vacuum Rabi splitting. The hybridisation leads to an avoided crossing, where the LP does not cross the exciton energy but strongly curves over as it approaches it. This leads to very strong dispersion and ultimately underlies the strongly reducing FSR for resonator modes at energies close to the exciton energy.

\begin{figure}[H]
    \centering
    \includegraphics[width=80mm]{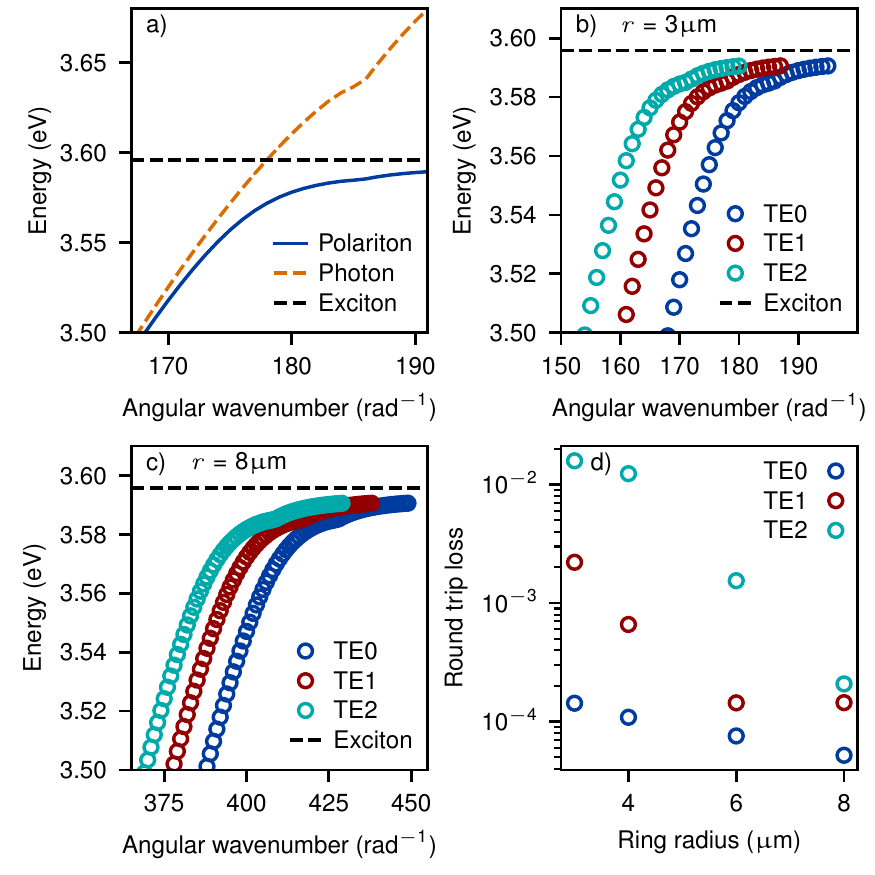}
    \caption{Dispersive properties of ring resonators computed for a system at 4 K. a) Dispersion of purely photonic (dashed orange) and polaritonic (solid blue) curved ridge waveguide with 3 $\mu$m radius of curvature.  b) Modes of a 3 $\mu$m polariton ring resonator. The first three transverse modes are shown. c) Modes of an 8 $\mu$m ring resonator. d) Round trip photonic loss of ring resonators vs. ring radius for the first three transverse modes. Losses are calculated for photons at the same energy as the exciton.}
    \label{fig:rr_disp}
\end{figure}

We now consider the dispersive properties of the ring resonators which we study in this work. Following a similar general method we first calculated the dispersion of purely photonic modes using electromagnetic simulation, and then calculated the dispersion of resonator polaritons using a coupled oscillator model. The photon dispersion was calculated using the finite difference eigenmode (FDE) solver of the commercial Lumerical MODE solutions package. This was used to simulate curved waveguides with transverse profile corresponding to the 2 $\mu$m wide ridge waveguides in the experiment. For a wide range of optical wavelengths the solver calculates the corresponding angular wavenumber $\hat{\beta}$ (rate of phase accumulation with angle travelled around the curved waveguide). The dashed orange curve in Fig.~\ref{fig:rr_disp}a) shows the photon energy vs. angular wavenumber. The polariton energy at each angular wavenumber is then obtained using the coupled oscillator model, as above, and is shown as the solid blue curve in Fig.~\ref{fig:rr_disp}a). The small kink in the curves around 3.64 eV for the photons and around 3.58 eV for polaritons is due to the rapid variation of the background refractive index of the quantum wells near the band edge. It is well above the energies of the experimentally investigated polariton modes and does not affect the results.

Once the dispersion of curved waveguides is known the energies of ring resonator modes can be calculated. The phase accumulated in one round trip of a circular resonator is $2\pi\hat{\beta}$ and the condition for resonance is that the round trip phase should be a multiple of $2\pi$. Thus we obtain $\hat{\beta}=n$ where $n$ is an integer. Interpolating into the curved waveguide dispersion gives the polariton energies of the resonances. These are shown in Figs.~\ref{fig:rr_disp}b) and c) for 3 $\mu$m and 8 $\mu$m rings, respectively. A feature of relatively wide resonators such at the 2 $\mu$m devices we use is that they support several transverse modes in addition to the longitudinal modes. This leads to a set of longitudinal modes for each transverse modal index. Figs.~\ref{fig:rr_disp}b) and c) show the first three transverse-electric polarised transverse mode families, TE0, TE1 and TE2. The transverse-magnetic polarised modes have a weak coupling to the quantum wells [1] and do not play a role in these experiments. The energy difference between any given longitudinal mode and the closest mode in the next transverse family is less than the distance between longitudinal modes.



\begin{figure}[H]
    \centering
    \includegraphics[width=\columnwidth]{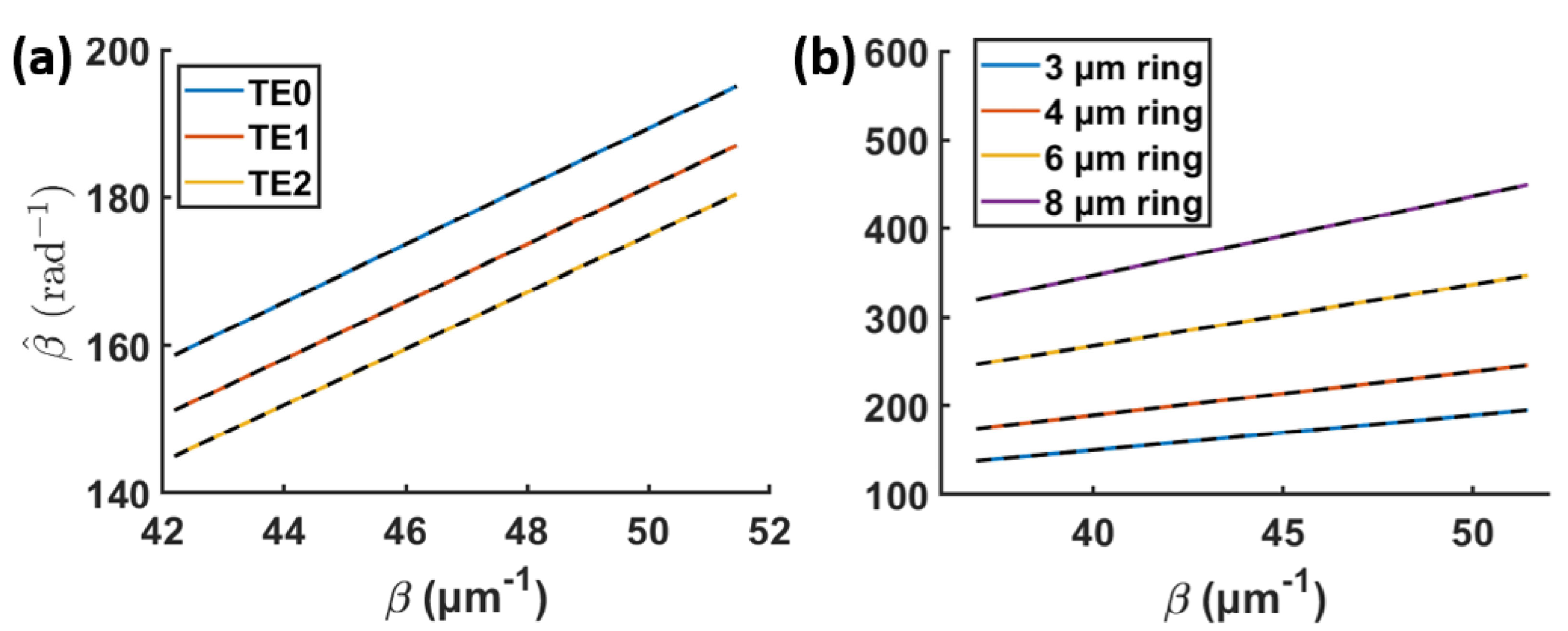}
    \caption{Computed angular wavenumber of the ring resonator modes vs. wavenumber of the planar waveguide from which the rings are etched. a) Values for the three lowest order TE modes of the 3 micron ring. b) Values for the lowest order TE mode of different rings. Coloured curves are the computed values. Dashed black lines are the best fit straight lines to each curve.}
    \label{fig:SI_reff}
\end{figure}

As discussed above we calculated the angular wavenumber of the pure photonic modes of the rings for a range of frequencies. At the same frequencies we also calculated the wavenumber of pure photons in the planar waveguide from which the rings were etched (also using Lumerical MODE). In Fig.~\ref{fig:SI_reff} we plot the angular wavenumbers $\hat{\beta}$ vs. the planar waveguide wavenumber $\beta$. Fig.~\ref{fig:SI_reff}(a) shows this for the three lowest order TE modes of the 3 micron ring. The coloured curves are the simulated data. The black dashed lines are best fit straight lines. It is seen that $\hat{\beta}$ is a linear function of $\beta$ to a high precision over a large range of wavenumbers, which correspond to photon wavelengths of 335 nm to 400 nm. This is true for all the different modes. Fig.~\ref{fig:SI_reff}(b) shows the same data for the lowest order TE mode of several different rings of different radii, showing the same linear relationship. Mathematically we have $\hat{\beta}_{m}(E)=\hat{\beta}_{m,0} + r_m\beta(E)$ where $\hat{\beta}_{m,0}$ is an offset analagous to the quantisation energy coming from the transverse confinement, and and $r_m$ is an effective radius of propagation for the mode labelled by index $m$. The radius comes in because we are comparing angular wavenumber (in units of rad$^{-1}$) and actual wavenumber (in units of m$^{-1}$). The offset $\hat{\beta}_{m,0}$ will disappear when taking derivatives to find, for example, the group velocity. The dispersive properties of all these modes (which have different orders and propagate in different rings) can therefore be explained entirely by the dispersion of the underlying planar waveguide. We conclude that the transverse confinement and circular propagation have negligible effect on the modal dispersive properties and that all transverse modes are equivalent to within a linear scaling factor related to the radius.

We finally consider the radiation losses in the resonators. Fig.~\ref{fig:rr_disp}d) shows the calculated round trip loss for photons in the resonator due to radiation loss and tunnelling through the cladding into the substrate. The round trip losses are obtained from the imaginary part of $\hat{\beta}$ as calculated by the FDE solver using perfectly-matched-layer boundary conditions. As expected, the losses are higher for the rings with smaller radius. The losses are also higher for higher order transverse modes TE1 and TE2 compared to the fundamental transverse mode family TE0. This suggests that either higher order transverse modes experience higher radiation losses or that their lower effective refractive index increases tunnelling loss through the substrate, or both.

\subsection{Frequency dependence of free spectral range as evidence of strong coupling}

A key signature of strong coupling is the anti-crossing of photon and exciton states which leads to a strong curvature of the lower polariton dispersion as it asymptotically approaches the exciton energy with increasing wavenumber (see section~\ref{sec:SI_dispersion}). A good measure of this curvature is the group velocity $v_g = \left(\partial E / \partial k\right)/\hbar$. For uncurved (linear) dispersion relations it is just a constant whereas in the strong coupling regime it is expected to vary strongly, starting from the pure-photon group velocity and tending to zero as the polariton energy approaches the exciton energy from below. In a simplified picture of a linear dispersion relation for the pure photons the polariton group velocity is proportional to the photonic fraction. In reality the pure photon dispersion relation also exhibits some curvature due to the wavelength dependence of the material refractive indexes, especially near the material band edges. As discussed in supplementary section~\ref{sec:SI_dispersion}, the dispersive properties (e.g. $v_g(E)$) of our ring resonator modes are determined entirely by those of the planar waveguide from which the rings are etched.

Experimentally we have direct access to the group velocity since it is directly proportional to the FSR. We have $v_g = \left(E_{\mathrm{FSR}}\cdot r\right)/\hbar$ where $r$ is a radius and $E_{\mathrm{FSR}}$ is the energy spacing between ring resonator modes (free spectral range). Because of this direct proportionality the variation of the FSR with energy is also a good measure of the polaritonic character of the modes. Theoretically, we can also calculate the expected free spectral range for both polaritons and also pure photons, as discussed in supplementary section~\ref{sec:SI_dispersion}.
Since we do not have experimental access to the transverse profile of the lasing modes, we calculate the FSR for a radius corresponding to the center of the 2 micron wide ridge for comparison with the experiment. The exciton energy, used in Eqn.~\ref{eq:coupled oscillator} to calculate the polariton energies, is taken from the spectra. The Rabi splitting is treated as a fitting parameter and we obtain values between 55 and 65 meV, in good agreement with the value reported in Ref. [1].

\begin{figure}[H]
    \centering
    \includegraphics[width=80mm]{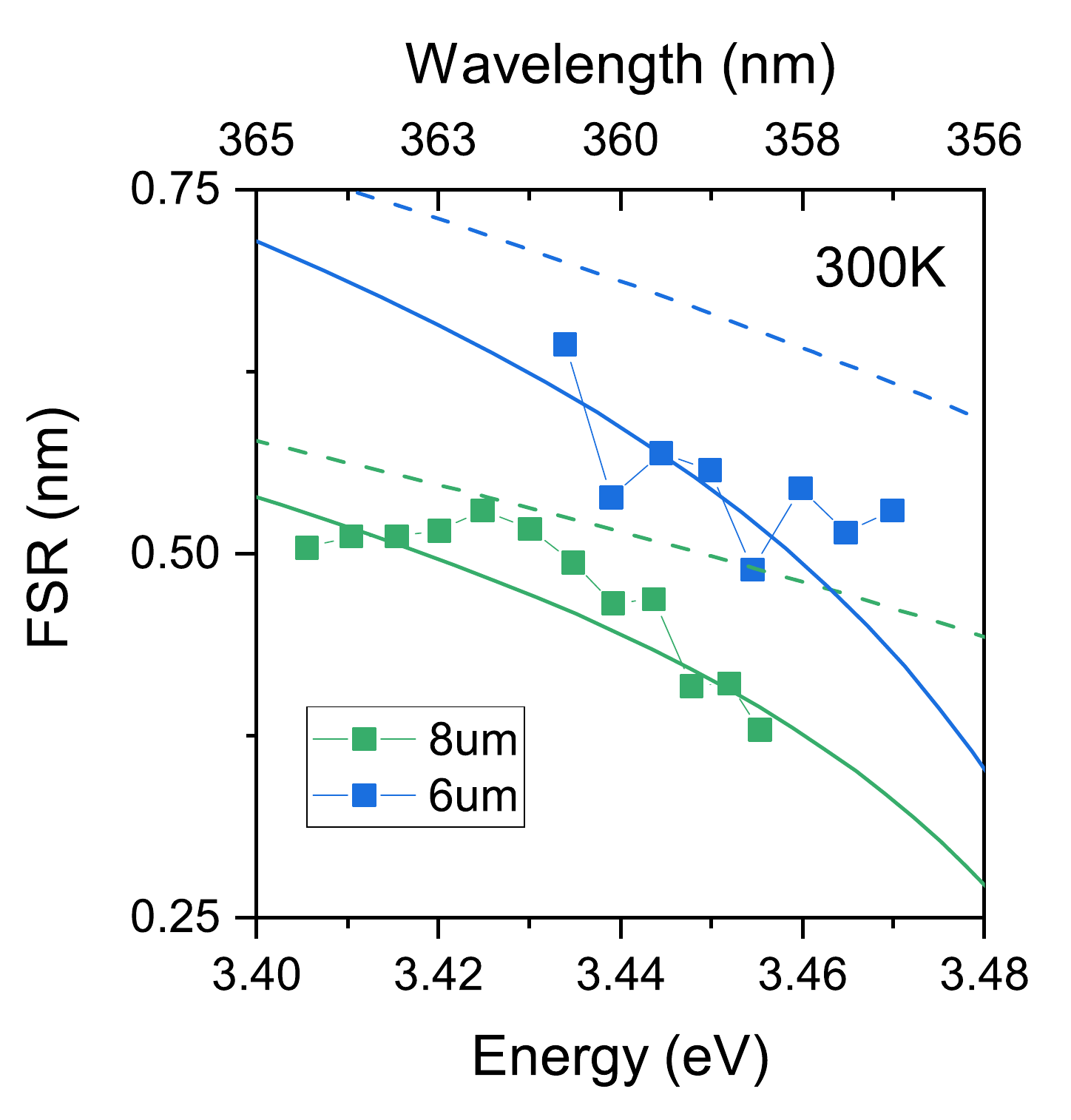}
    \caption{FSR of microring resonators measured for radii of 6 and 8 $\mu$m), with a 2 $\mu$m width, at pulse energy above the polariton lasing threshold (1.7 nJ) taken at 300 K. The solid line represents a theoretical model accounting for polariton non-linearities, whereas the dashed line represents a purely photonic model.}
    \label{fig:fsr300k}
\end{figure}

Figure 2b in the main text shows the experimentally measured FSR for several rings at 4 K. In the case of the 3 $\mu$m radius ring the experimental FSR (and hence the group velocity) decreases by 65\% of the maximum measured value over an energy range of 41 meV. As discussed in the main text this large variation in FSR cannot be accounted for by a purely photonic model. The rings with other radii at 4 K show similar behaviour, with changes in FSR (group velocity) too large for the purely photonic model to explain. They evidence the strong coupling and its strong effect on the curvature of the dispersion relation.

In Fig.~\ref{fig:fsr300k} we show the experimental FSR for two rings at 300 K as points. For the 8 $\mu$m radius ring the FSR varies by 25\% between lowest and highest energy points over a 54 meV range of energies. This is less than the case at 4 K since the data points at 300 K are for energies further from the exciton where the polaritons are more photonic and the curvature due to strong coupling is therefore expected to be lower. The solid (dashed) curves show the theoretical FSR for polaritons (pure photons). For 8 $\mu$m radius rings the theory predicts a variation of 24\% for polaritons over the same energy range as the experimental points. The purely photonic model can account for a change of only 14\%. 
Overall at 300 K there is better agreement with the strongly coupled model than the purely photonic one, which confirms that the strong coupling is retained in the lasing regime up to room temperature. This is further supported by our observation that above threshold the luminescence peak from the exciton does not exhibit broadening or energy blueshift, which would be expected if we were beyond the Mott density where strong coupling collapses (see main text).

\subsection{PL spectra without G\lowercase{a}N background}
At 4 K  the polariton emission intensity was integrated in the region where the contribution from the QW and GaN exciton emissions is negligible.

\par
\begin{figure}[H]
    \centering
    \includegraphics[width=80mm]{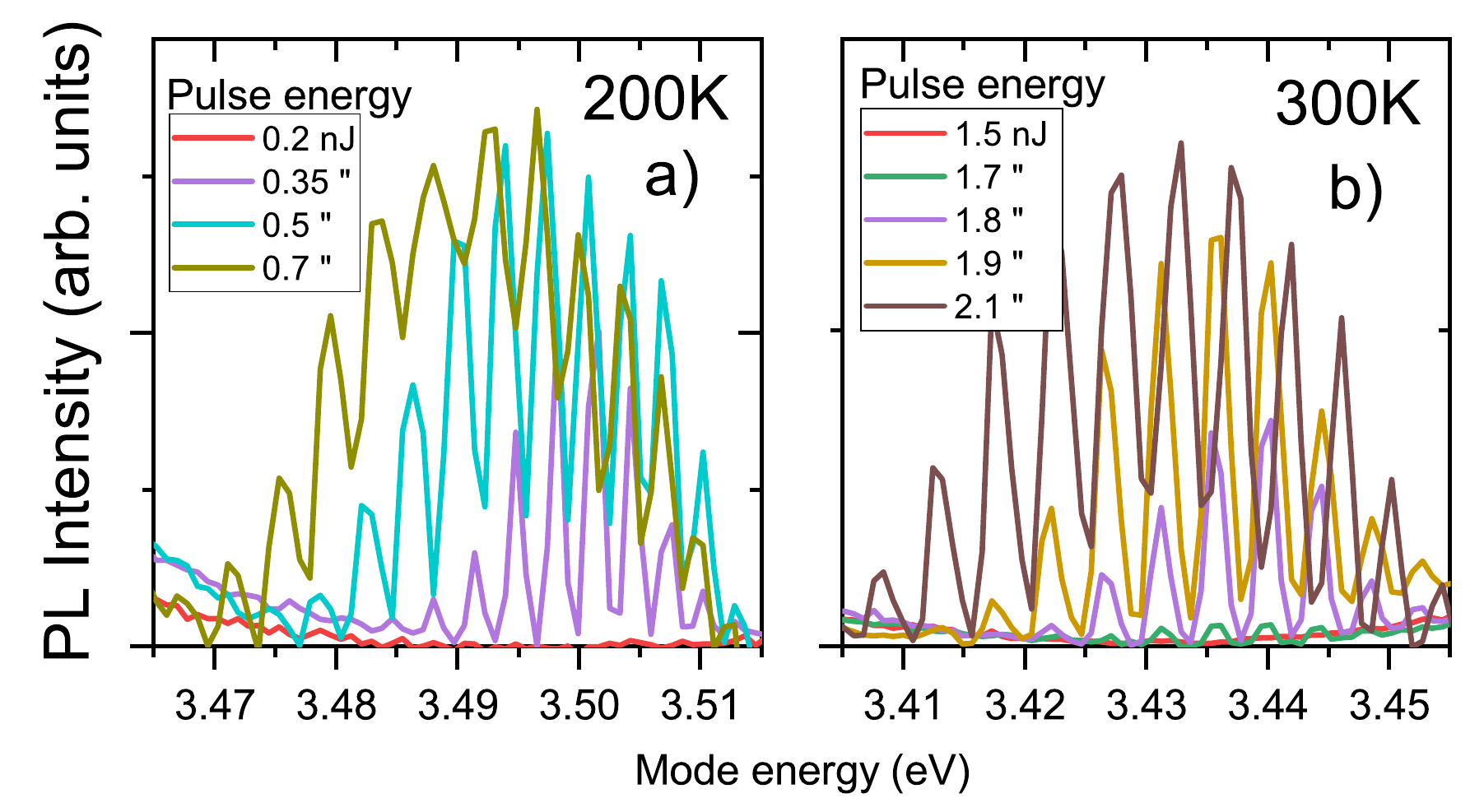}
    \caption{PL spectra of the 8 $\mu$m ring resonator recorded at $T$ = 200, and 300 K, with the PL background resulting from the GaN exciton peaks removed. The bottom line of the graph is zero intensity.}
    \label{fig:pl_no_gan}
\end{figure}

By contrast, the polariton modes at 200 and 300 K sit on a strong, smooth, and broad incoherent PL background coming mainly from the bulk GaN excitons. This background was removed by fitting it with a Gaussian peak. In the main text, our analysis to extract the threshold-like behaviour of the modes relies on the removal of this background for enhanced contrast. In Figs. \ref{fig:pl_no_gan}.a and b, we present the PL spectra measured at 200 and 300 K of the 8 $\mu$m ring with the PL background removed. These spectra show how the modes quickly grow in intensity, much faster than the linear evolution of the GaN exciton peaks themselves (cf. next SM section).

\subsection{Power dependency of the bulk G\lowercase{a}N excitonic luminescence peaks and QW  exciton luminescence peaks at different temperatures}

\begin{figure}[H]
    \centering
    \includegraphics[width=80mm]{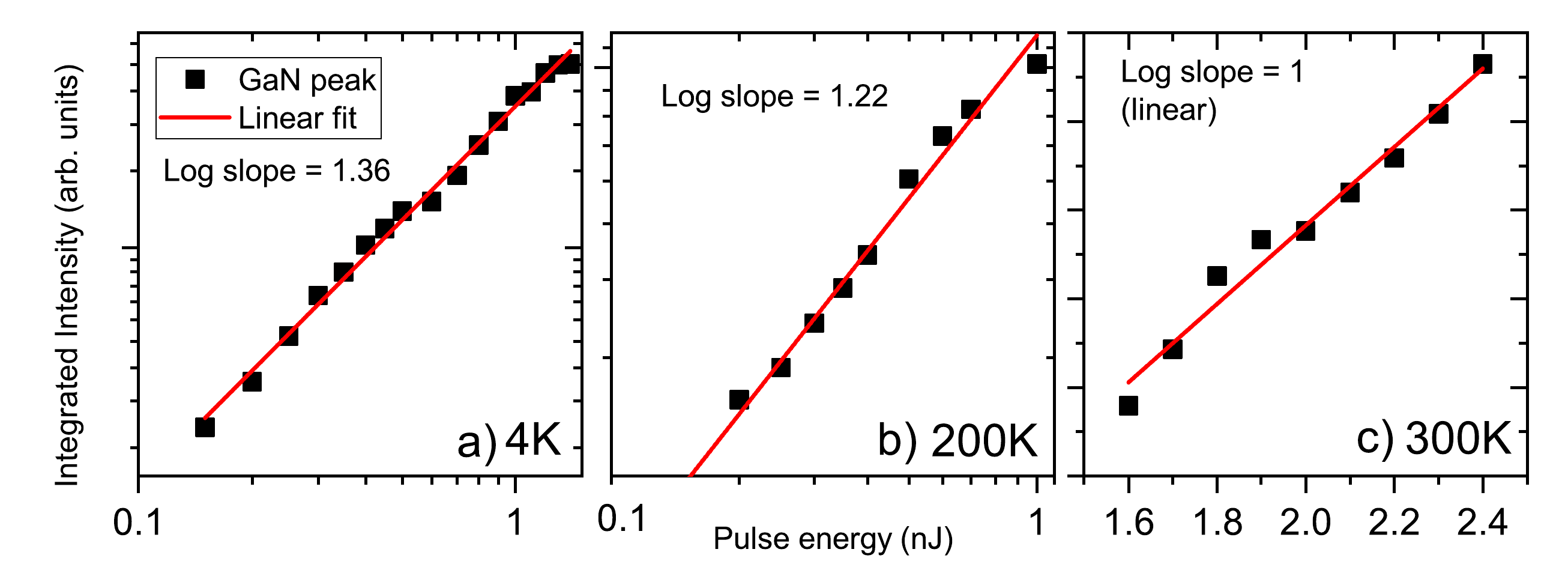}
    \caption{Integrated PL intensity of the bulk GaN exciton peaks recorded at 4, 200, and 300 K for the 8 $\mu$m ring. The red solid lines represent quasi-linear fits with the corresponding slope indicated in each figure. Log-log scale at 4 and 200 K. Linear scale at 300 K.}
    \label{fig:gan_8}
\end{figure}

One argument used to prove polariton lasing operation is the presence of a sharp threshold at which the PL intensity of the lasing modes rapidly increases. In this section, we give the pulse energy dependency of the integrated PL intensity of the bulk GaN exciton peaks at 4, 200, and 300 K for the 8 $\mu$m microring resonator (Fig. \ref{fig:gan_8}).

\begin{figure}[H]
    \centering
    \includegraphics[width=80mm]{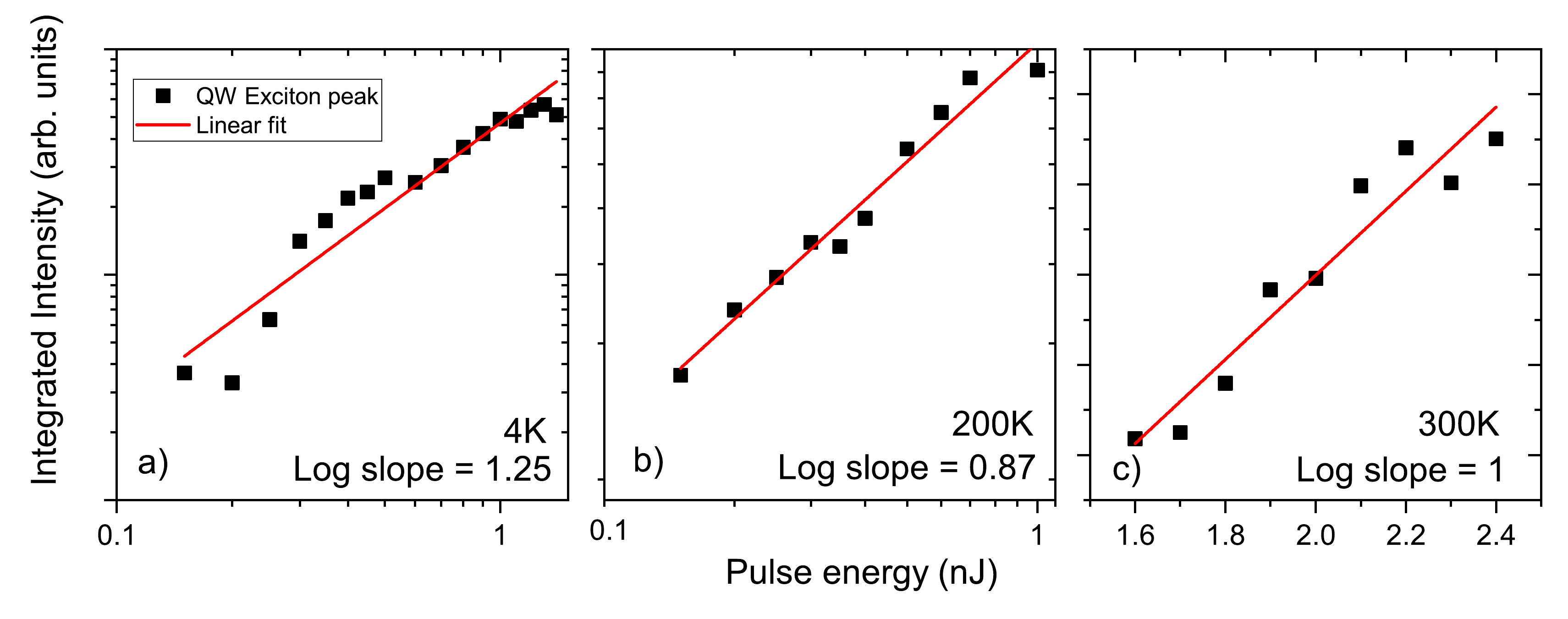}
    \caption{Integrated PL intensity of the QW excitonic peaks at 4, 200, and 300 K for the 8 $\mu$m ring. The red solid lines represent fits with the corresponding slope indicated in each figure. Doubly logarithmic scale for 4 and 200 K. Linear scale for 300 K.}
    \label{fig:exciton_integ_intensity}
\end{figure}

Note that from the log-log scale used to display the data recorded at 4 and 200 K, a slope slightly higher than unity is obtained for the power dependency of the integrated PL intensity of the bulk GaN exciton peaks, which implies a weak nonlinearity. For example, at 4 K, we have a fit leading to $y \sim x^{1.36}$. All the exponents remain reasonably close to 1 and hence the power dependence of the bulk GaN exciton peak PL intensity is found to be close to linear. We use a linear scale at 300 K due to the smaller accessible range of powers. This is in sharp contrast with the polariton modes, whose intensity follows a superlinear increase well-above a quadratic function (see the main text). The same analysis applies for the QW excitonic peaks as seen in Fig. \ref{fig:exciton_integ_intensity}. Similar results were obtained for the 4 $\mu$m microring and are shown in Fig. 1 of the main text. 

\subsection{Blueshift of the 4 $\mu$m ring}

\begin{figure}[H]
    \centering
    \includegraphics[width=80mm]{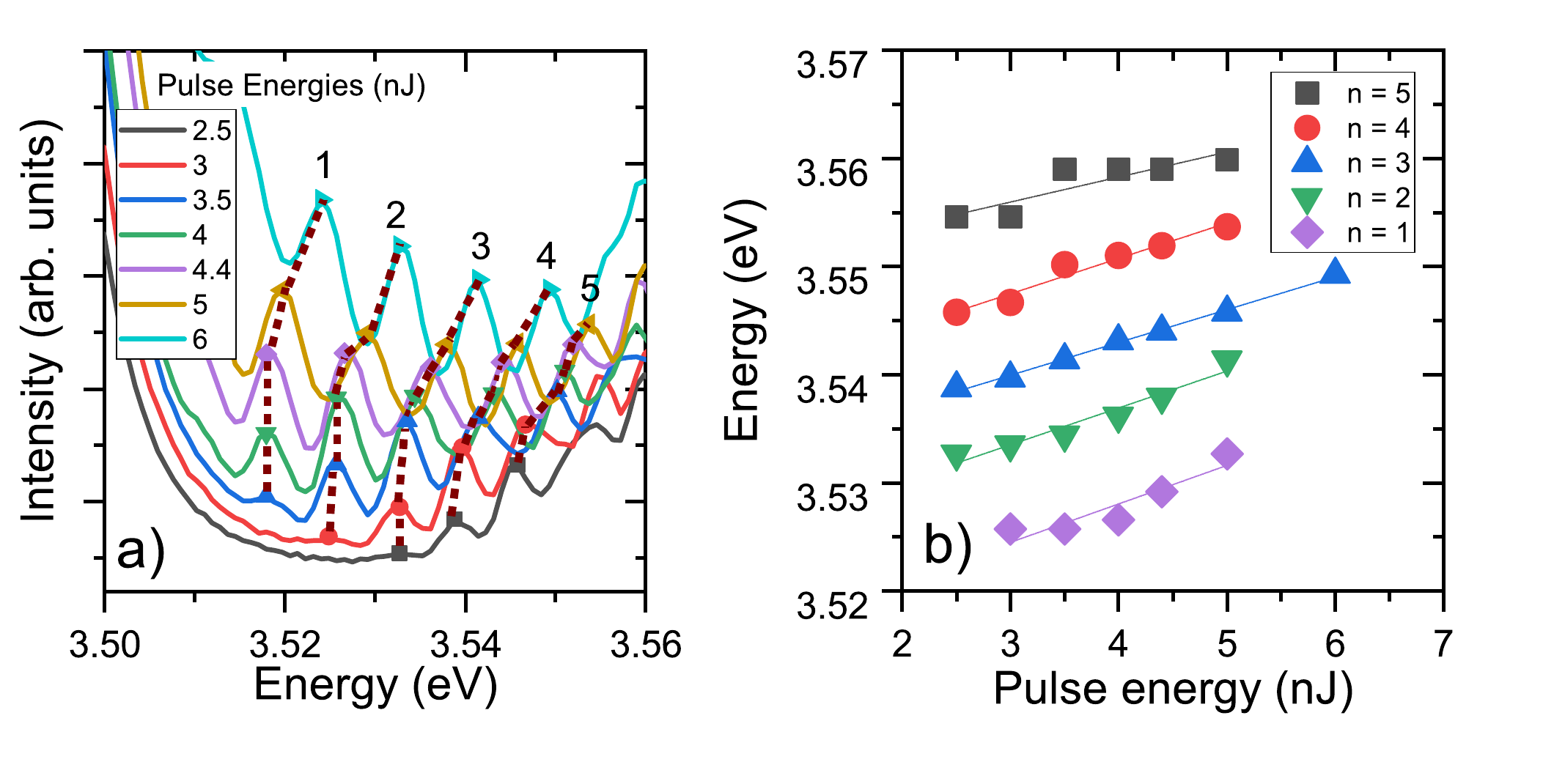}
    \caption{a) PL spectra of the 4 $\mu$m ring recorded at 4 K at different pulse energies, above threshold, showing the blueshift of the modes. The lines are guides for the eyes. b) Evolution of the peak positions as a function of pulse energy. The solid line is a linear fit.}
    \label{fig:4um_blueshift}
\end{figure}

Blueshift is a defining feature of polaritons. It is a signature of their non-linearity owing to their excitonic part. In the main text, we have described strong blueshift in the 8 $\mu$m ring at all temperatures. Such blueshift is also seen in the smaller 4 $\mu$m ring, as shown in Fig. \ref{fig:4um_blueshift}. The difference in the range of pulse energies is due to a slightly different excitation scheme compared to the main text, exciting a smaller part of the ring with each pulse. Under this excitation scheme, the blueshift is much clearer than in Figure 1 of the main text, but happens at higher pulse powers.

[1] J. Ciers, J. G. Roch, J.-F. Carlin, G. Jacopin, R. Butt{\'{e}}, and N. Grandjean,
  Propagating Polaritons in {III}-Nitride Slab Waveguides, Phys. Rev. Applied \textbf{7},
  034019 (2017).\\
	
[2] D. M. Di Paola, P. M. Walker, R. P. A. Emmanuele, A. V. Yulin, J. Ciers, Z. Zaidi, J.-F. Carlin, N. Grandjean, I. Shelykh, M. S. Skolnick, R. Butt{\'{e}}, and D. N. Krizhanovskii,
Ultrafast-nonlinear ultraviolet pulse modulation in an {AlInGaN} polariton waveguide operating up to room temperature, Nat. Commun. \textbf{12}, 3504 (2021).\\

[3] Stephen Thoms and Douglas S. Macintyre, Investigation of {CSAR} 62,  a new resist for electron beam lithography, J. Vac. Sci. Technol. B \textbf{32}, 06FJ01 (2014).\\

[4] A. L. Schawlow and C. H. Townes, Infrared and optical masers, Phys. Rev. \textbf{112}, 1940--1949 (1958).\\

\end{document}